\documentclass[aps,pre,twocolumn,showpacs,groupedaddress,amsmath,amssymb,floatfix]{revtex4}
\usepackage{graphicx}
\usepackage{epstopdf}

\begin{document}


\title{Shear Thickening and Scaling of the Elastic Modulus in a Fractal Colloidal System with Attractive Interactions}


\author{Chinedum O. Osuji}
\email[]{chinedum.osuji@yale.edu}
\thanks{Present Address: Department of Chemical Engineering, Yale University,
New Haven CT 06511}
\affiliation{School of Engineering and Applied Sciences, Harvard
University Cambridge MA 02138}
\author{Chanjoong Kim}
\affiliation{School of Engineering and Applied Sciences, Harvard
University Cambridge MA 02138}.
\author{David A. Weitz}
\affiliation{School of Engineering and Applied Sciences, Harvard
University Cambridge MA 02138}


\date{\today}

\begin{abstract}
Dilute oil dispersions of fractal carbon black particles with attractive Van
der Waals interactions display continuous shear thickening followed
by shear thinning at high shear rates. The shear thickening
transition occurs at  $\dot\gamma_{c}\approx 10^{2}-10^{3}s^{-1}$
and is driven by hydrodynamic breakup of clusters. Pre-shearing
dispersions at shear rates $\dot\gamma>\dot\gamma_{c}$ produces
enhanced-modulus gels where $G' \sim \sigma_{pre-shear}^{1.5-2}$ and
is directly proportional to the residual stress in the gel measured
at a fixed sample age. The observed data can be accounted for using
a simple scaling model for the breakup of fractal clusters under
shear stress.
\end{abstract}


\maketitle


Colloidal particles interacting with attractive potentials in a
fluid form gels above some critical volume fraction, $\phi_{c}$,
where $\phi_{c}$ is a function of the interaction strength, $U$. The
viscoelastic properties of the gels are determined by $\phi$, $U$
and the topology of the network of particle contacts, characterized
by a fractal dimension, $d_{f}$. Additionally, the properties are
sensitive to the mechanical history of the system. To obtain
reproducible results, rheological studies of colloidal gels
typically employ a high rate pre-shear as a rejuvenation or
initialization step in which the mechanical history of the gel is
effectively erased. The non-Newtonian flow curves associated with
attractive colloidal suspensions typically display monotonic shear
thinning and as a result the modulus established after rejuvenation
is not overly sensitive to the exact pre-shear flow condition. Shear
thickening transitions are neither generally known nor predicted for
systems where attractive interactions are sufficiently strong to
induce flocculated gels \cite{COO:Zukoski2004_1,COO:Raghavan1997_1}.
Such transitions would be accompanied by drastic changes in the
underlying fluid microstructure. As a result, the elasticity of gels
would show a marked dependence on the nature of the pre-shear flow
and its location on the flow curve relative to the shear thickening
regime.

Here we report on the observation of shear thickening for attractive
colloidal particles in a simple Newtonian fluid. We find the
elasticity of gels formed by pre-shearing above the shear thickening
transition scales as a power law with the pre-shear stress, and is
directly proportional to the internal stress in the sample, which we
measure. We propose a scaling model which considers the dependence
of the gel modulus on the cluster number density produced during
pre-shear flow. The model shows good agreement with the experimental
data.

Measurements were made at 25$^{\circ}$C on dispersions of
carbon black particles ranging from 2 to 8 wt.\%.  A non-polar,
small-molecule fluid, tetradecane, was used as the suspending medium
to avoid complicating influences of electrostatics or adsorption
of macromolecular solvent species onto the colloidal surface. The
carbon black (Cabot Vulcan XC72R) exists as $\approx 0.5 \mu m$
diameter fumed particles with fractal dimension $d_{f}^{p} = 2.2$
and $\rho_{carbon}$ = 1.8 g/cm$^{3}$ . Tetradecane ($\rho_{solvent}$
= 0.762 g/cm$^{3}$, $\eta_{solvent}$ = 2.8 mPa.s) was obtained from
Aldrich Chem. Co.. Samples were prepared by vigorous mixing
of the two components using a vortexer. Optical studies were
conducted using a Bohlin CS  rheometer with a transparent base-plate
through which samples were imaged using a CCD camera.
Rheological measurements were made in strain-control mode
using a TA Instruments AR-G2 rheometer with cone-plate and
double-wall Couette geometries. The fast hardware feedback of this
instrument, when run in strain-control, enabled measurements of the
internal stress within samples by applying continuously computed
torques to maintain constant tool position. Samples were pre-sheared
before each measurement to remove the effect of shear history using
an empirically determined shear rate (100s$^{-1}$) which provided consistent
results \cite{COO:TrappeWeitz_Scaling_1}. Samples were checked using
various geometries (standard and roughened) to ensure the
absence of rheological artifacts and phenomena such as wall slip and
edge-fracture. They were also replicated using a pure
strain-controlled instrument, ARES-LS1 (Rheometrics).

\begin{figure}[ht]
\includegraphics[width=65mm, scale=1]{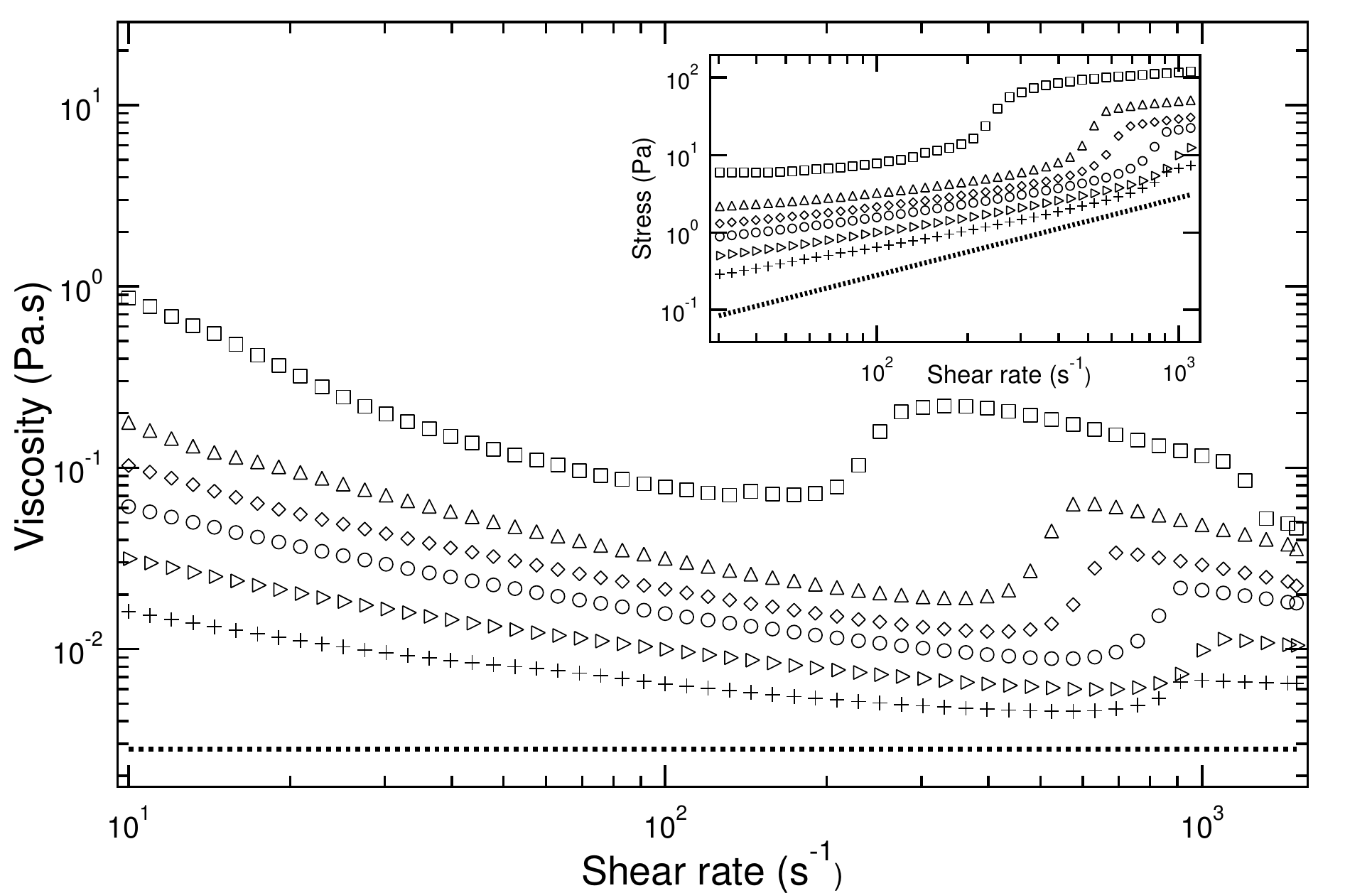}
\caption{Shear rate dependent viscosity for a series of weight
fractions. $+:2\%$, $\triangleright:3\%$, $\bigcirc:4\%$,
$\diamond:5\%$, $\triangle:6\%$, $\square:8\%$ The dotted-line is
the background viscosity of the solvent. Inset: Rate dependent shear
stress.} \label{flow_curves}
\end{figure}

\begin{figure}[h]
\includegraphics[width=55mm]{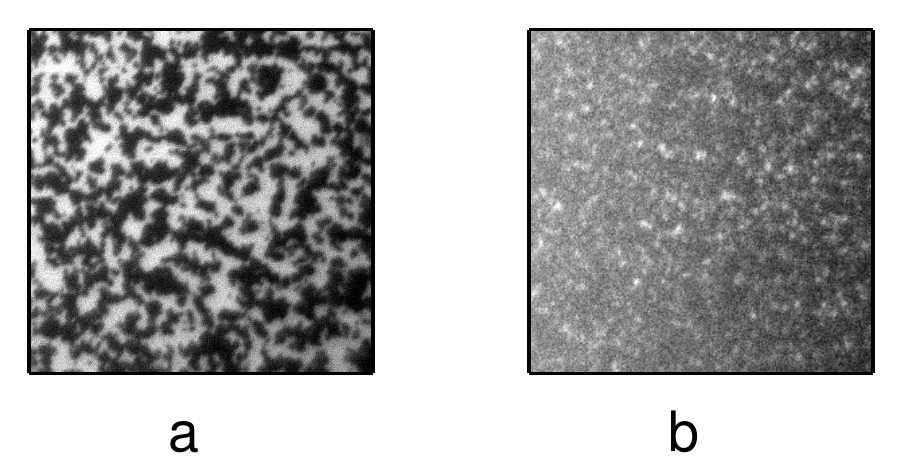}
\caption{Microstructure under shear in parallel plate geometry. Gap
= $100 \mu m$, $\phi= 3$ wt.\%, 1.5x1.5 mm area. (a) $\dot\gamma =
133s^{-1}$ (b) $\dot\gamma = 1330s^{-1}$. Illumination for (b) was
several times higher than for (a) to obtain sufficient light
transmission through the optically dense structure. Pixel binning
was used to decrease the required exposure time.}
\label{optical_rheology}
\end{figure}

We measure the steady state viscosity as a function of shear rate
using a cone-plate geometry.  At low shear rates, between $\approx
10^{0}$ and $10^{1}s^{-1}$ optical studies using the plate-plate
geometry show a build-up of structure in the sample via the
formation of rolling, vorticity aligned cylindrical flocs. Such
shear-induced structures have been observed in other viscoelastic
systems under flow such as thixotropic clay gels
\cite{COO:Pignon_PRL1997}, nanotube suspensions
\cite{COO:Hobbie_PRL2004} and attractive emulsion droplets
\cite{COO:Pasquali_PRL2004}. Viscometry of such ``sticky'' systems
at low shear rates may be susceptible to wall slip effects
\cite{walls2003ysa} and so caution is required in interpreting any
rheological data in this regime. At higher shear rates where
homogeneous flows are readily achieved, $\dot\gamma>10 s^{-1}$,
Figure \ref{flow_curves}, there is modest shear thinning accompanied
by the formation of densified clusters as shown in Figure
\ref{optical_rheology}a. Shear thinning persists until $\approx
10^{2}-10^{3}s^{-1}$ where there is a composition dependent
continuous transition to a shear thickening flow. The flow curves
exhibit a hysteresis loop on descending sweeps as is typical of
thixotropic materials, and show good reproducibility on subsequent
ascending-descending loops. Shear thickening has in fact been
reported \cite{COO:Kato2001_1}, though not extensively considered,
for a system consisting of carbon black particles in an adsorbing
silicone oil. This was attributed to hydro-cluster formation due to
the occurrence of stress overshooting at shear rates that produced
shear thickening. In the present case, at high shear rates in the
shear thickening regime, the system presents a finely dispersed
microstructure in which the densified clusters produced at lower
shear rates have been eroded. This results in a higher effective
volume fraction of particles in the dispersion, and thus an enhanced
viscosity, Figure \ref{optical_rheology}b. In contrast to the hard
sphere case, shear thickening results not from hydro-cluster
formation, but from the breakup of locally dense clusters of the
fractal colloidal particles into less dense structures which
increase viscous dissipation in the system. A negative thixotropy of
ferric-oxide suspensions composed of acicular particles has also
been attributed to a qualitatively similar mechanism
\cite{kanai1995ntf}.

Based on the dispersed particle size, we estimate the P\'eclet
number, $Pe=\frac{\sigma a^{3}}{k_{B}T}\approx 10^{2}-10^{3}$,
confirming that the high shear rate regime is dominated by
hydrodynamic forces. Similarly, the particle Reynold's number
$Re=\frac{\rho \dot\gamma a^{2}}{\eta}\approx 10^{-4}-10^{-5}$,
indicating that viscous effects dominate and contributions from
inertial forces should be minimal. Under similar conditions in
Stokesian dynamics simulations of hard spheres
\cite{COO:Melrose1996_1,COO:SinghNott2000_1} a transition is
observed from positive to negative normal stresses on shear
thickening due to hydrocluster formation at high $Pe$, with $|N_1|
\sim \dot\gamma^{1}$ where $N_1$ is the first normal stress. In the
present system, inertia corrected $N_1$ transitioned to slight net
positive values at the onset of shear thickening
\cite{COO:Osuji_Weitz_SoftMatter2008}, suggesting, along with the
optical data, that hydrocluster formation is not in fact the
mechanism in operation here. The critical shear rate at the onset of
shear thickening, $\dot\gamma_{c}$, scales roughly as $\phi^{\beta}$
and the critical stress $\sigma_{c}$ as $\phi^{\nu}$ with $\beta$
roughly between -1.5 and -2, and $\nu\approx 1$. From measurements
performed using high viscosity basestock oil ($\eta \approx$ 33
mPa.s), we observe that the critical shear rate is inversely
proportional to the solvent viscosity, suggesting that shear
thickening in this system is controlled by a critical stress, rather
than by a critical rate.

\begin{figure}[h]
\includegraphics[width=65mm, scale=1]{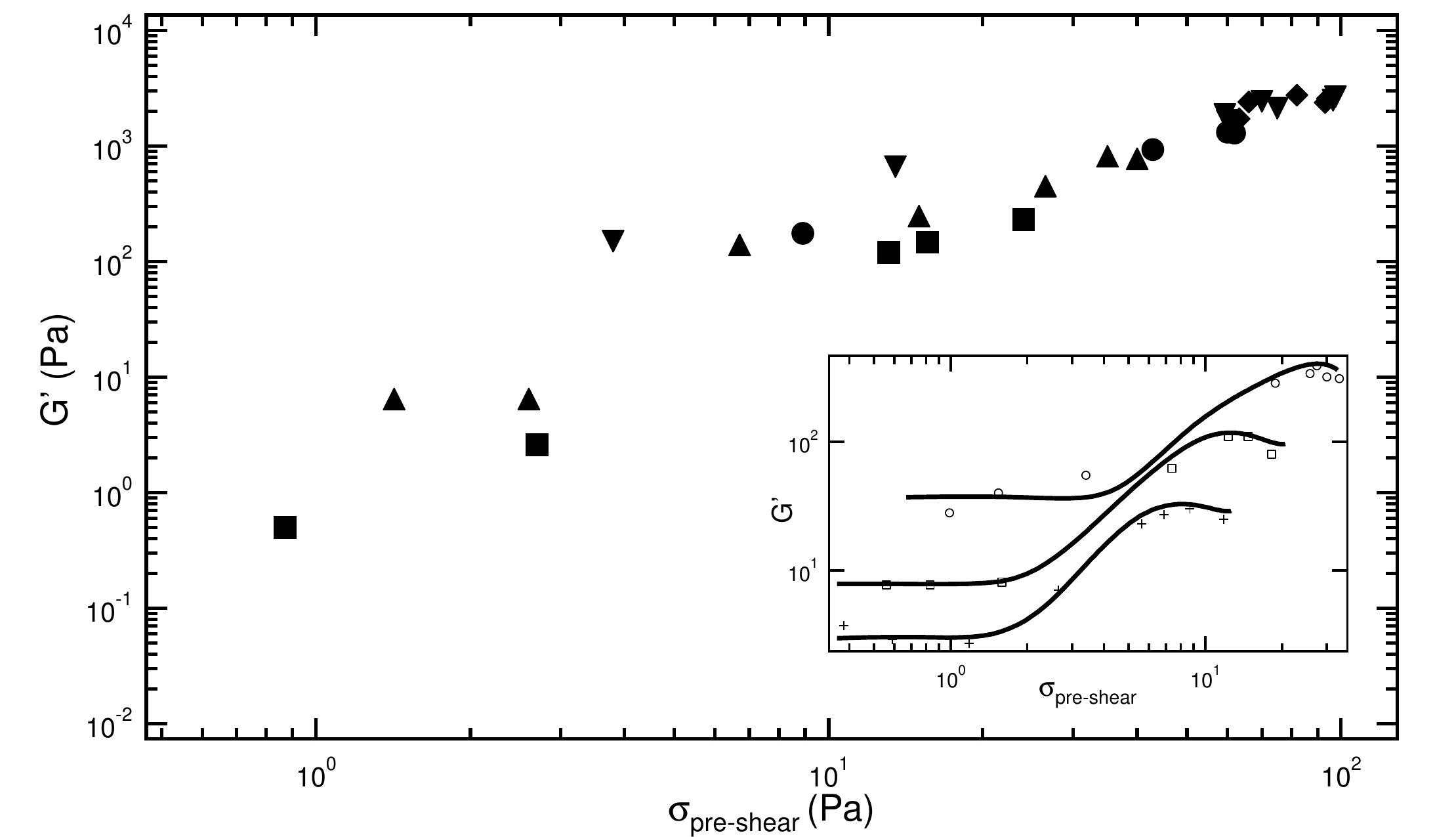}
\caption{Elasticity of shear thickened gels as a function of the
pre-shear stress, using cone-plate geometry. $\blacksquare:3\%$,
$\blacktriangle:4\%$, {\large $\bullet$}$:5\%$, $\blacklozenge:6\%$,
$\blacktriangledown:8\%$. Inset: Data obtained spanning low and high
stresses using a double-wall Couette cell. $+:2\%$, $\square:3\%$,
$\bigcirc:5\%$} \label{modulus_shear_stress}
\end{figure}

We use dynamic measurements to study the dependence of gel modulus
on pre-shear flow and composition. Samples were sheared to remove
the effects of flow history and then pre-sheared at the rate of
interest for 20 minutes, more than sufficient to achieve a steady
flow and constant viscosity. The system was then allowed to sit
quiescently for 30 minutes. We measure the viscoelastic storage and
loss moduli, $G'$ and $G''$, in the linear regime, $\gamma=0.05\%$
at a fixed angular frequency $\omega=1$\,rad/s., where $\gamma$ is
the shear strain. The elastic modulus of gels formed by pre-shearing
at a fixed rate in the shear thickening regime follows a typical
power law $G' \sim \phi^{\alpha}$, with exponent $\alpha=3.5$, as
might be expected for attractive systems
\cite{COO:Grant_Russel_phi_scaling}. In the shear-thinning regime,
before the thickening transition, the exponent is slightly higher,
$\alpha=4.2$. The elasticity of thickened gels is strongly enhanced,
typically by over one order of magnitude, compared to gels produced
by non shear-thickening flow. This suggests that in addition to
volume fraction, $\phi$, there is another parameter dependence
involved in determining the elasticity of gels produced across a
range of shear rates. The modulus follows the same qualitative
dependence on shear rate as the viscosity, with a sudden upturn at
the critical rate or critical shear stress. A double-wall Couette
cell was used to mitigate the effects of sedimentation at low shear
stresses where samples did not form robust thickened gels, or were
too dilute to be gravitationally stable for study in the cone-plate
geometry. Remarkably, as in Figure
\ref{modulus_shear_stress}(inset), the dependence of the gel modulus
on the pre-shear stress has a  common form across different
compositions, $\phi$, such that the data can be scaled onto a single
curve using the critical stress, $\sigma_c$, and critical modulus,
$G'_c$, as scaling parameters. Data of shear thickened samples show
a striking dependence of the modulus on the pre-shear stress, $G'
\sim\sigma^{1.5-2}$, as shown in Figure
\ref{modulus_shear_stress}(main). These samples which were studied
in the cone-plate geometry at large pre-shear stresses where the
resulting high modulus gels were not sensitive to sedimentation.
There is a somewhat weaker volume fraction dependence such that in
the shear thickening regime, overall, the modulus of the colloidal
gel is substantially influenced by the pre-shear stress applied to
the system prior to gelation.

We interpret these results in terms of the dependence of the cluster
size on shear stress. As the stress increases, densified clusters
break, and assume a size that is set by the balance between the
shear stress and the cohesive energy density of the cluster.

The shear force on the cluster, $F_{shear} = 4\pi\sigma R_{c}^2$
where $R_{c}$ is the radius of the fractal cluster. This force is
spread across $N_{s}$ particles in the plane of shear of the cluster
where $N_{s}\sim R_{c}^{d_{f}-1}$ where $d_{f}$ is the fractal
dimension of the cluster. This force is balanced by the attractive
interaction between particles, defined by the interaction strength
$U$ and the relevant length scale, $a$, the particle diameter.
Thus we have

\begin{equation}
R_{c}^{d_{f}-3} \sim \frac {4\pi\sigma a}{U}
\label{eq:cluster_size_scaling}
\end{equation}

which sets the dependence of the equilibrium cluster size on the
shear stress. The elastic modulus scales directly with the number
density of clusters, i.e. $G'\sim\nu\, U$, the cohesive energy
density, as in many disordered systems\cite{COO:Larson_text}. The
number density, $\nu$ of clusters of radius $R_{c}$ scales as
$\phi/R_{c}^{d_{f}}$ so that overall, we expect the gel modulus to
scale as

\begin{equation}
G' \sim \phi \sigma^{\frac{d_f}{3-d_f}}
\label{eq:final_modulus_scaling}
\end{equation}

\begin{figure}[h]
\includegraphics[width=65mm, scale=1]{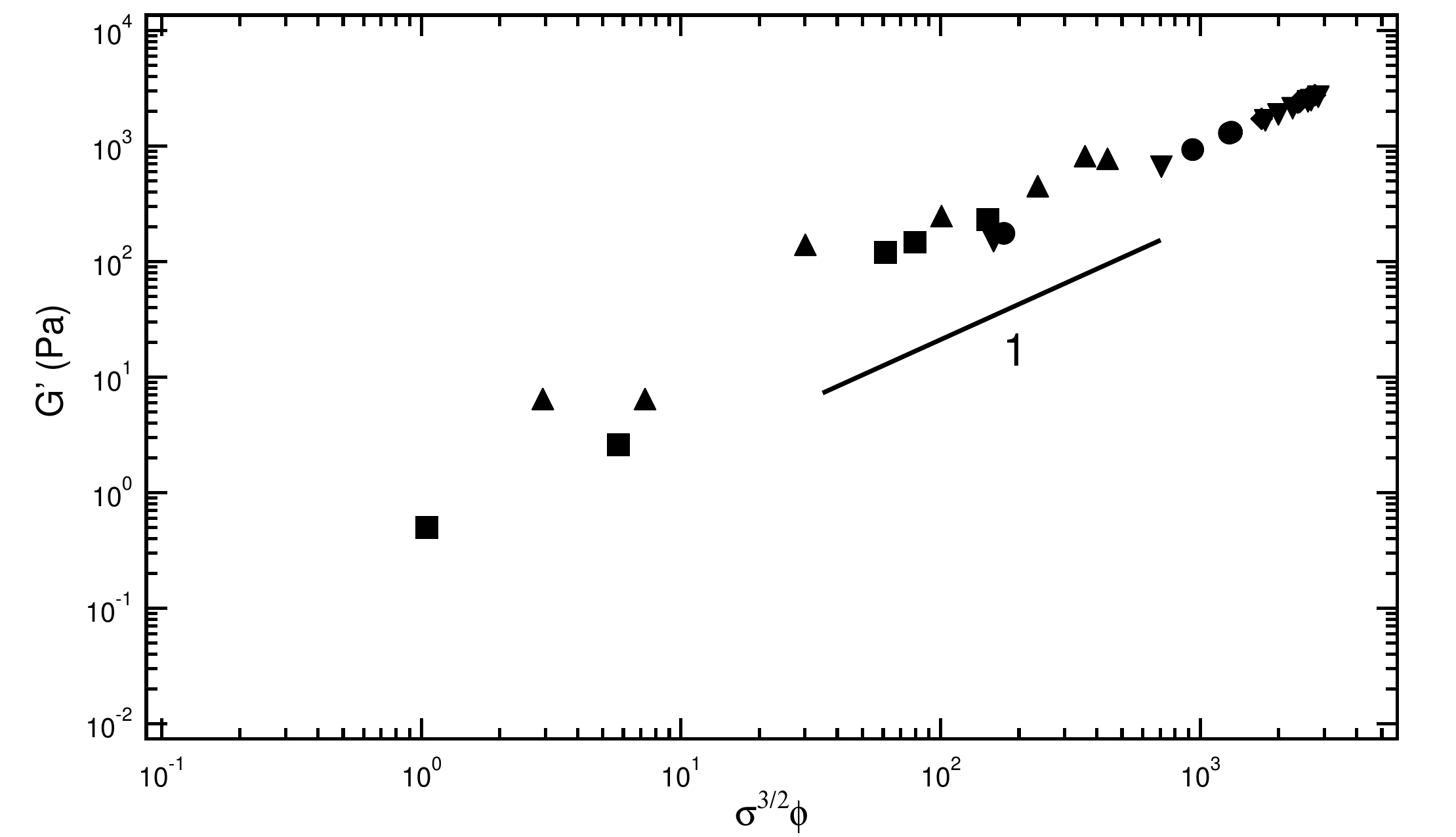}
\caption{Elasticity of shear thickened gels as a function of the
pre-shear stress re-scaled according to equation
\eqref{eq:final_modulus_scaling} with $d_{f}=1.8$.
$\blacksquare:3\%$, $\blacktriangle:4\%$, {\large $\bullet$}$:5\%$,
$\blacklozenge:6\%$, $\blacktriangledown:8\%$. The solid line is a
visual guide with a slope of 1.} \label{scaled_modulus_shear_stress}
\end{figure}

Assuming a cluster fractal dimension of 1.8, as is found in
diffusion limited aggregating systems then $G'
\sim\sigma^{3/2}\phi$. We plot this in Figure
\ref{scaled_modulus_shear_stress} and find good accord between our
data and a slope of 1 as shown, over several decades. Our scaling is
in rough agreement with empirical data on flocculated systems where
the number of particles in a floc scales roughly as
$N_{c}\sim\sigma^{-1}$ \cite{COO:SonntagRussel1987_Experiment}.
Although the simple assumptions of the scaling argument provide
satisfactory results, the choice for $d_f$ should be rationalized by
more detailed measurements than we have so far made.

\begin{figure}[ht]
\includegraphics[width=65mm, scale=1]{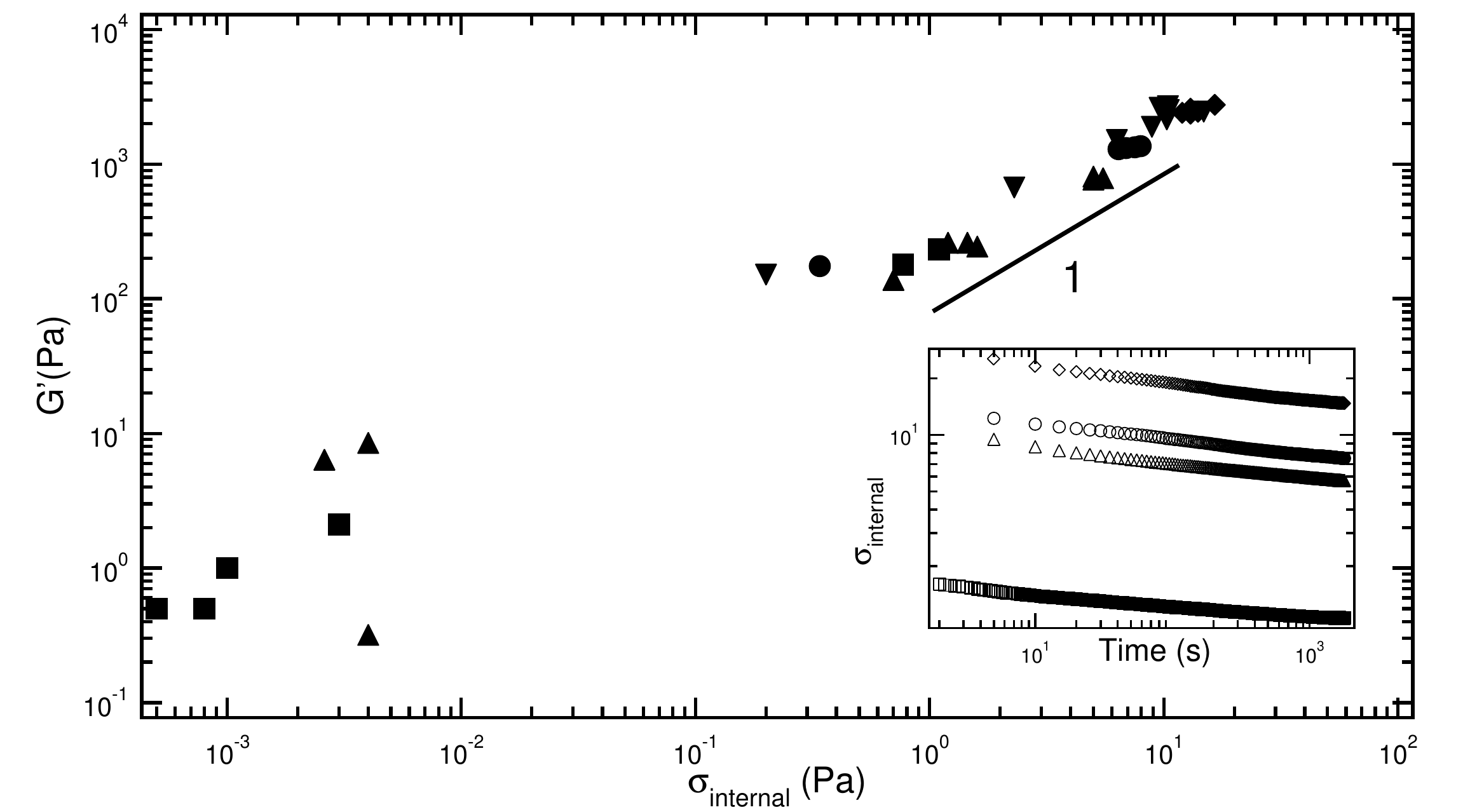}
\caption{Shear modulus scales linearly with residual stresses
resulting from the quench into the gel state on cessation of flow.
$\blacksquare:3\%$, $\blacktriangle:4\%$, {\large $\bullet$}$:5\%$,
$\blacklozenge:6\%$, $\blacktriangledown:8\%$. The solid line is a
visual guide with a slope of 1. Inset: Time dependence of internal
stress, with open symbols for clarity.
\label{modulus_internal_stress}}
\end{figure}

The abrupt cessation of shear in a freely flowing system above
$\phi_{c}$ results in a sudden entry into the gel state. The system
loses mobility as particles experience sticky contacts with
neighbors leading to the formation of structures that rapidly span
space, forming an elastic medium. This rapid sol-gel transition
should be characterized by the development of internal or residual
stresses, $\sigma_{i}$, coincident with the mechanical quench - the
network structure, deformed with respect to an equilibrium relaxed
configuration, would exhibit stresses which relax gradually in time.
These stresses are believed to drive the peculiar dynamics observed
in the aging of some glassy systems
\cite{COO:Bandhyopadhyay_Mochrie_Leheny,COO:Bellour_Munch}. Since
they result from a deformation of the structure, the internal
stresses are expected to be proportional to the elastic modulus
\cite{COO:Cipelletti2005_slow_dynamics_1}, though no direct
experimental evidence as yet confirms this. Previous work has
examined the very short time ($t\lesssim 0.1s$) dynamics of stresses
in weakly flocculating systems subjected to stress jumps
\cite{COO:Mewis2005_1}. Here, we followed $\sigma_{i}(t)$ in our
system from $t\approx 1 s$ after cessation of flow to $t\approx$ 30
mins. and find a weak power law decay over this period. For shear
thickened gels, the stress scales roughly as $\sigma_{i}\sim
t^{-0.1}$, inset Figure \ref{modulus_internal_stress}. The internal
stress acts counter to the direction in which the sample was
pre-sheared and is reversed on changing the pre-shear direction,
indicating that the deformation of the clusters during flow
(relative to the quiescent state) gives rise to the residual stress
that is manifested by the rate quench and rapid gelation.
Critically, we find that the internal stress at a fixed sample age,
$t$=30 mins., is directly proportional to the modulus of the system
measured immediately thereafter, over several orders of magnitude of
internal stress, produced by different pre-shear flows, Figure
\ref{modulus_internal_stress}. The prototypical picture of a slow
diffusion limited gelation clearly does not apply despite the
relatively low volume fractions under study here. Gelation is
unusually rapid and the system displays a well developed elastic
modulus at the shortest times studied, $\approx$ 1 second after
cessation of shear.

Shear thickening in hard sphere and repulsive Brownian systems has
been rationalized using a force balance model that compares the
shear force pushing particles together to the diffusive flux that
inhibits cluster formation \cite{COO:Wagner2001_1}.

\begin{equation}
\frac{3\pi\mu\dot\gamma_{c}a^{3}}{2h}= -k_{B}T\frac{\partial \ln
g(r)}{\partial r}
\end{equation}

where $a$ is the particle radius, $h=r-2a$ the distance between two
particles, $g(r)$ the particle distribution function and $\mu$ the
viscosity of the medium, replaceable by the suspension viscosity,
$\eta$, as a way of accommodating many body effects. In the system
under study here, at low shear stresses, large non-Brownian clusters
resulting from breakage of the gel at low shear rates have no
diffusive barrier towards further clustering. They densify as shear
stresses force them into contact, reducing their effective volume
fraction in the system. Above a critical stress however, the
cohesive energy density of the clusters is overcome and they start
to disintegrate with a attendant rise in viscosity.

The observation of shear thickening in this system of attractive
particles was surprising. Indeed, previous careful examinations of
flocculated systems have not resulted in observation of shear
thickening in fractal \cite{COO:Raghavan1997_1} and ``solid''
particulate systems \cite{COO:Zukoski2004_1}. A particularly
systematic examination of the role of interaction strength using
depletion forces \cite{COO:Zukoski2004_1} shows that shear
thickening diminished and eventually disappeared on addition of
substantial attractive interactions. It is clear that in the current
system, the fractal nature of the particles, irreversibly fused
agglomerates of smaller particles produced by flame pyrolysis, plays
a key role. It drives the increase in the effective volume fraction
on cluster break-up. Shear thickening is thus a product of shear
induced reduction of cluster size and increase of the effective
volume fraction of the particles in the dispersion due to the
fractal particle structure. Correspondingly, the modulus of shear
thickened dispersions is enhanced due to the increase in cluster
number density. The scaling of the elastic modulus with pre-shear
stress is satisfactorily accounted for by consideration of the
equilibrium cluster size as a function of this stress. The direct
proportionality between the elastic modulus and the internal stress
resulting from a system quench on cessation of shear has been
confirmed by direct measurement. The dynamics of these internal
stresses are a topic for continued studies.

\begin{acknowledgments}
The authors gratefully acknowledge L. Cipelletti, V. Trappe and H.
Wyss for  helpful discussions, A. Negi for supplementary rheological
measurements and the Infineum Corporation for funding.
\end{acknowledgments}

\bibliography{colloidal_rheology}

\end{document}